\documentclass[conference]{IEEEtran}
\usepackage{cite}
\usepackage{amsmath,amssymb,amsfonts}
\usepackage{algorithmic}
\usepackage{graphicx}
\usepackage{textcomp}
\usepackage{xcolor}
\usepackage{booktabs}
\usepackage{tabularx}
\usepackage{ragged2e}
\usepackage{array}
\usepackage{url}
\usepackage{caption}
\usepackage{makecell}
\usepackage{amsmath}
\usepackage{amssymb}
\usepackage{float}
\usepackage{mathtools}
\usepackage{algorithm}
\usepackage{algorithmic}
\usepackage{multirow} 
\usepackage{siunitx}
\usepackage[caption=false,font=footnotesize]{subfig}
\usepackage{placeins}
\usepackage[colorlinks=true,linkcolor=black,citecolor=black,urlcolor=black]{hyperref}

\def\BibTeX{{\rm B\kern-.05em{\sc i\kern-.025em b}\kern-.08em
    T\kern-.1667em\lower.7ex\hbox{E}\kern-.125emX}}

\newcommand{\blfootnote}[1]{%
  \begingroup
  \renewcommand\thefootnote{}\footnote{#1}%
  \addtocounter{footnote}{-1}%
  \endgroup
}
\begin{document}

\title{Adaptive Split-MMD Training for Small-Sample Cross-Dataset P300 EEG Classification}
\thanks{This work was supported by NSF Grant No. 2423943.}

\author{\IEEEauthorblockN{Weiyu Chen}
\IEEEauthorblockA{\textit{Swartz Center for Computational Neuroscience,} \\\textit{Institute for Neural Computation} \\
\textit{University of California San Diego}\\
La Jolla, USA \\
\textit{The Department of Electronic and Electrical Engineering},\\
\textit{Southern University of Science and Technology} \\
\textit{Shenzhen, Guangdong 518055, China} \\
12210460@mail.sustech.edu.cn}
\and
\IEEEauthorblockN{Arnaud Delorme}
\IEEEauthorblockA{\textit{Swartz Center for Computational Neuroscience,} \\\textit{Institute for Neural Computation} \\
\textit{University of California San Diego}\\
La Jolla, USA \\
\textit{Centre de recherche Cerveau et Cognition} \\
\textit{Paul Sabatier University}\\
Toulouse, France \\
arnodelorme@gmail.com}
}

\maketitle
\begingroup
\hypersetup{urlcolor=blue}%
\blfootnote{Code available in the ``P3 transfer learning'' example on \url{https://eegdash.org}.}
\endgroup

\begin{abstract}
Detecting single-trial P300 from EEG is difficult when only a few labeled trials are available. When attempting to boost a small target set with a large source dataset through transfer learning, cross-dataset shift arises. To address this challenge, we study transfer between two public visual-oddball ERP datasets using five shared electrodes (Fz, Pz, P3, P4, Oz) under a strict small-sample regime (target: 10 trials/subject; source: 80 trials/subject). We introduce Adaptive Split Maximum Mean Discrepancy Training (AS-MMD), which combines (i) a target-weighted loss with warm-up tied to the square root of the source/target size ratio, (ii) Split Batch Normalization (Split-BN) with shared affine parameters and per-domain running statistics, and (iii) a parameter-free logit-level Radial Basis Function kernel Maximum Mean Discrepancy (RBF-MMD) term using the median-bandwidth heuristic. Implemented on an EEG Conformer, AS-MMD is backbone-agnostic and leaves the inference-time model unchanged. Across both transfer directions, it outperforms target-only and pooled training (Active Visual Oddball: accuracy/AUC 0.66/0.74; ERP CORE P3: 0.61/0.65), with gains over pooling significant under corrected paired t-tests. Ablations attribute improvements to all three components.
\end{abstract}


\section{Introduction}
\label{sec:intro}
Event-related potentials (ERPs) provide rapid, noninvasive readouts of cognition. Among them, the P300 (P3) elicited by oddball tasks underpins applications in attention monitoring and target detection, which is a cornerstone of many BCI and cognitive assessment pipelines.
However, due to time constraints, fatigue, or clinical limitations, difficulties have arisen in collecting large volumes of data. Therefore, cross-dataset learning becomes attractive by leveraging existing large datasets to help a new, small deployment dataset. The core challenge is distribution shift across labs, devices and montages, protocols, cohorts, and noise, which breaks the i.i.d.\ assumptions of standard training. These domain shifts also make small-sample adaptation hard to stabilize \cite{HeWu2020EA}. 

A practical operationalization of cross-dataset learning is to use both a large labeled source and a small target dataset jointly while explicitly handling the inter-dataset shift. A typical strategy is to pre-train on a larger source dataset and fine-tune on the small target set, or to naively pool source and target trials. However, naive pooling is known to bias the learner toward the source distribution \cite{Zhao2019,Xu2021}, while fine-tuning on only a few trials is brittle and highly sensitive to normalization drift \cite{Jayaram2016,Kostas2020ThinkerInvariance,HeWu2020EA}.

Motivated by these limitations, we adopt a joint-training strategy and introduce a minimal modification to standard supervised learning that addresses three pain points: domain size imbalance, normalization drift, and dataset shift. We refer to this joint recipe as Adaptive Split–MMD Training (AS-MMD). First, a target-focused loss weight counteracts source dominance without changing sampling ratios; the weight follows a warm-up schedule and scales with the square-root of the source/target size ratio, clipped to a stable range. Second, Split-BN keeps a single set of affine parameters while maintaining separate running statistics per domain; training alternates domains with snapshot/restore of buffers, and inference on the target domain always uses the target buffers. Third, an RBF-MMD penalty on logits (with median-bandwidth heuristic) gently aligns source and target decision spaces without introducing extra trainable components. We instantiate models with an EEG Conformer to balance expressivity and efficiency under a modest parameter budget; the recipe is backbone-agnostic and does not alter the inference-time architecture \cite{Song2023EEGConformer}.

Concretely, we study transfer between two public visual-oddball ERP datasets \cite{Isbell2025CESCA,Kappenman2021ERPCORE}. Following prior studies that identify the P300 as most prominent over centro-parietal and occipital regions \cite{Polich2007P3Review,AlvaradoGonzalez2016P300}, we restrict analysis to the five electrodes that are both shared across the two datasets and most P3-relevant (Fz, Pz, P3, P4, and Oz). We enforce a strict small-sample regime with 10 trials per subject on the deployment dataset and 80 per subject on the source. Across both transfer directions, we compare against target-only training and naive pooling under a 5$\times$5 cross-validation protocol.

The main contributions of this paper are as follows:
\begin{itemize}
\item We empirically demonstrate that co-training on a large source dataset together with a small target dataset improves the target dataset’s single-trial P3 accuracy compared with target-only training and naive pooling, under a strict small-sample protocol (10 trials/subject) and 5$\times$5 cross-validation in both transfer directions.
\item We propose an innovation in training, named Adaptive Split-MMD Training (AS-MMD), which integrates target-focused weighting, Split-BN, and parameter-free MMD alignment into a backbone-agnostic recipe, yielding consistent target-domain gains without adding new trainable modules or altering the inference-time architecture. 
\end{itemize}

\section{Related Work}
Cross-subject and cross-dataset generalization in ERP/P300 decoding has been approached from both shallow and deep perspectives. Compact CNNs tailored to EEG (e.g., EEGNet with depthwise separable convolutions) established strong baselines across paradigms, while deeper ConvNets with visualization demonstrated end-to-end decoding from raw EEG \cite{Lawhern2018EEGNet,Schirrmeister2017HBM}. Transformer-style backbones that incorporate self-attention have also been explored for ERP/P300 to capture longer-range temporal dependencies; EEG Conformer is a representative compact convolution–attention hybrid for EEG decoding \cite{Song2023EEGConformer}. However, systematic evaluations have shown that distribution shifts across datasets can substantially degrade deep models, motivating explicit alignment and normalization strategies \cite{Xu2020CrossDataset}. 

Geometry-aware alignment has provided lightweight transfer mechanisms without expanding model capacity. Riemannian alignment (RA) matches second-order statistics on the SPD manifold prior to classification, and Euclidean alignment (EA) performs data-space normalization to reduce inter-subject variability; these methods remain competitive baselines and often serve as inexpensive front-ends to learned models \cite{Zanini2018RA,HeWu2020EA}.

In deep domain adaptation (DA), discrepancy-based objectives and normalization have been particularly impactful. Multi-kernel Maximum Mean Discrepancy (MMD) in Deep Adaptation Networks aligns hidden-layer distributions with minimal architectural overhead. Orthogonal to loss design, Adaptive Batch Normalization (AdaBN) replaces batch statistics using target data at test time, while Domain-Specific BN (DSBN) maintains per-domain buffers under shared weights, delivering robust adaptation without duplicating backbones \cite{Li2018AdaBN,Chang2019DSBN}. Beyond these, thinker-invariant training and source-sample selection specifically for P300 have further reduced negative transfer and improved cross-person performance in low-resource settings \cite{Kostas2020ThinkerInvariance,Kilani2024SourceSel}.
Deep Adaptation Networks (DAN) place MMD penalties at several hidden layers—typically the last task-specific layers—so that source/target representations are progressively aligned \cite{Long2015DAN}. In contrast, our recipe applies a single, parameter-free MMD at the \emph{logit} level to minimize added hyperparameters and computation in a few-shot ERP setting.

Positioned within this landscape, our approach combines (i) a target-prioritized loss schedule to guard against source dominance under severe size imbalance, (ii) split BN buffers (shared affine, per-domain statistics) for stable normalization across domains, and (iii) a parameter-free, logit-level RBF–MMD penalty to gently align decision spaces. The design is backbone-agnostic and adds no trainable modules, complementing prior DA and normalization techniques while directly addressing the small-target cross-dataset ERP setting. Our instantiation with EEG Conformer leverages its compact convolution–attention hybrid to capture both local and long-range structure in EEG while remaining efficient for low-channel configurations \cite{Song2023EEGConformer}.

\section{Datasets and Task}
\label{sec:datasets}

\subsection{Dataset}
\subsubsection{Dataset~1: Active Visual Oddball subset of ``Cognitive Electrophysiology in Socioeconomic Context in Adulthood''}
Dataset~1 originates from a public EEG collection of 127 young adults (18–30 years) curated in Brain Imaging Data Structure (BIDS) format with multiple ERP tasks derived from or adapted to ERP CORE paradigms. We restrict usage to the Active Visual Oddball (AVO) task to align with the P3b setting in our study. Although the cohort includes 127 participants, we selected 40 AVO participants with sufficient oddball trials. The included BIDS subject IDs were: s01--s10, s12--s14, s16--s17, s19--s24, s26, s28--s38, s40--s43, s47--s48, s51. This matches the 40-subject size of Dataset~2 for comparable folds and ensures feasible within-subject balanced sampling when drawing 80 trials per subject (40 oddball and 40 standard). The dataset description and access information are provided in the project documentation and OpenNeuro release \cite{Isbell2025CESCA}.

\subsubsection{Dataset~2: ERP CORE P3}
Dataset~2 is the P3 component of ERP CORE, a curated resource comprising optimized paradigms, experiment scripts, processing pipelines, and example data from 40 neurotypical adults. We use the P3 active visual oddball task and associated event schema. ERP CORE’s documentation and archival materials (Open Science Framework and article) provide comprehensive task details and analysis pipelines \cite{Kappenman2021ERPCORE}.

\subsection{Task Setup}
\label{subsec:task}

\subsubsection{Objective, label space, and event mapping}
We perform binary single-trial classification of oddball versus standard under an active visual oddball (P3b) paradigm. Labels are harmonized to a common schema (oddball = 1, standard = 0) using each dataset’s event annotations and mapping rules described below. Both datasets provide stimulus-onset event codes. We adopt a two-digit convention $XY$ where $X$ encodes the block’s designated target (A–E) and $Y$ encodes the current trial’s stimulus (A–E). A trial is labeled oddball if and only if the stimulus matches the block target, i.e., the diagonal codes $\{11, 22, 33, 44, 55\}$; all other stimulus codes are labeled standard. This rule mirrors the oddball logic in the ERP CORE P3 event specification and is applied uniformly to both datasets \cite{Kappenman2021ERPCORE}. We use only stimulus-onset events for labeling and ignore response or feedback markers.

\subsubsection{Common electrodes and preprocessing}
To minimize channel mismatch while retaining P3b sensitivity, we restrict analysis to the five common posterior–midline electrodes: Fz, Pz, P3, P4, and Oz, since it has been previously reported to be the most relevant to detect the P300 component \cite{AlvaradoGonzalez2016P300}. Raw EEG is harmonized by a unified pipeline: resampling to 128~Hz; epoching from $-100$~ms to $+1000$~ms relative to stimulus onset; band-pass filtering at 0.5–30~Hz with power-line notch; baseline correction using the pre-stimulus interval; and independent component analysis (ICA) performed on the full set of recorded channels. Artifact-related components (e.g., ocular and muscle) were identified following standard MNE/ERP CORE practice based on their characteristic scalp maps, time courses, and spectra, then removed before back-projecting the remaining components \cite{Gramfort2013MNEPython,Gramfort2014MNE,Jung2000ICA}. The cleaned data were subsequently restricted to the five common electrodes. ERP CORE provides reference processing scripts that motivated these settings and ensured consistent event timing \cite{Kappenman2021ERPCORE}.

\subsubsection{Evaluation protocol}
We evaluate two source–target configurations: (i) Dataset~1 as source with Dataset~2 as target; and (ii) Dataset~2 as source with Dataset~1 as target. For each configuration, we sample per subject 80 source trials and 10 target trials using stratified sampling. For the source set, we fix the oddball budget at 40 (the within-set minimum) and draw 40 standard trials at random to achieve a 1:1 balance—i.e., 40/40—obtained by downsampling standards from the native $\sim$1:4 (oddball:standard) ratio. For the target set, we enforce a strict 5/5 oddball–standard split (10 trials total) per subject, mirroring the source-set balancing. Cross-validation uses 5 folds and 5 random seeds (5$\times$5). This protocol standardizes per-subject class priors and reduces variance due to imbalance, while preserving the cross-dataset distribution shift.

\section{Method}
\label{sec:method}

\subsection{Backbone}
\label{subsec:backbone}
We use an EEG Conformer backbone implemented via the Braindecode Python library \cite{Schirrmeister2017HBM}. EEG Conformer is a compact convolutional--Transformer architecture designed for EEG decoding \cite{Song2023EEGConformer}. It combines convolutional layers to extract local temporal--spatial patterns with Transformer self-attention to capture long-range temporal and inter-channel dependencies. Concretely, the network comprises:
(i) a temporal/spatial convolutional front end that reduces sequence length and projects raw EEG into embeddings;
(ii) sinusoidal positional encoding to inject temporal order;
(iii) a stack of $L=3$ Transformer encoder layers, each containing multi-head self-attention with $H=10$ heads, a feed-forward sublayer of dimension $4d$, dropout (0.1), Gaussian Error Linear Unit (GELU) activations, residual connections, and LayerNorm;
and (iv) a classifier head that applies average pooling over temporal tokens followed by a linear projection to logits $z=f(x;\theta)\in\mathbb{R}^{C}$ with $C{=}2$ classes (oddball vs.\ standard).
We use embedding dimension $d{=}40$. This architecture is compact (only $\sim$200k parameters in our setting) yet expressive enough to model the spatio-temporal structure of single-trial ERP signals. All implementations follow the Braindecode toolbox.

\subsection{Baselines}
\label{subsec:baselines}
Both baselines use the same backbone, optimizer, regularization, and early-stopping criteria as Sec.~\ref{subsec:train}; they differ only in training data composition and the absence of domain-specific mechanisms.

\subsubsection{Target-only}
Training uses only the target dataset. The loss is standard cross-entropy on target batches; Batch Normalization (BN) keeps a single set of running statistics; no alignment loss is applied. This baseline isolates performance under strict small-sample conditions.

\subsubsection{Pooled (Source + Target)}
Training uses the union of source and target datasets with cross-entropy, a single set of BN running statistics pooled across domains, and no alignment term. Sampling follows the per-subject trial budgets defined in Sec.~\ref{sec:datasets}; no explicit domain weighting is applied.

These baselines reflect standard supervised pipelines commonly used in EEG decoding without domain adaptation or domain-aware normalization.

\subsection{AS-MMD}
\label{subsec:as_mmd}
 AS-MMD combines a target-weighted supervised loss, a logit-level RBF-MMD alignment term, and Split-BN with shared affine parameters and per-domain running buffers. 
\begin{figure*}[t]
    \centering
    \includegraphics[width=\textwidth]{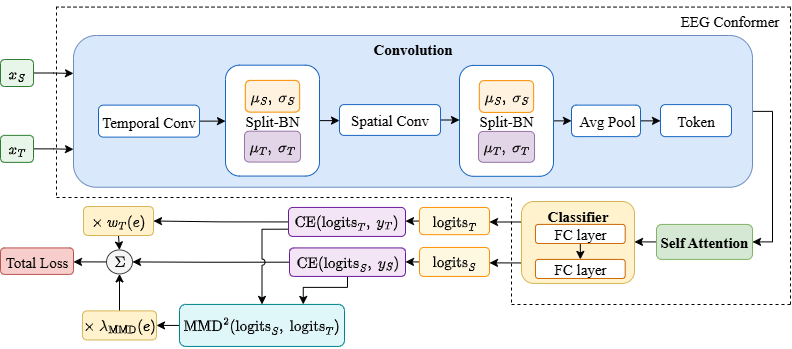}
    \caption{Overview of AS-MMD on the EEG Conformer backbone.}
    \label{fig:method}
\end{figure*}

The overall training pipeline is shown in Fig.~\ref{fig:method}.
Source and target mini-batches are processed by the same EEG Conformer.
Within the backbone, data pass sequentially through a temporal convolution, a Split-BN layer with domain-specific running statistics, a spatial convolution, a second Split-BN, and average pooling to form a single token representation. 
This token is fed to self-attention and a small two-layer classifier, producing logits for the source and target streams.
At the head, we compute cross-entropy on both streams, with the target term up-weighted over training by $w_T(e)$. 
In parallel, we impose a logit-level RBF-MMD between the two sets of logits, scaled by $\lambda_{\mathrm{MMD}}(e)$. 
The supervised terms and the alignment penalty are then summed to yield the total loss.

\vspace{1em}
\noindent{Symbols shown in Fig.~\ref{fig:method}}:
\begin{itemize}
\item $x_S, x_T$: source/target input trials.
\item $\mu_S,\sigma_S^2$ and $\mu_T,\sigma_T^2$: Split-BN running mean/variance (source vs.\ target).
\item $\mathrm{logits}_S, \mathrm{logits}_T$: classifier logits for source/target branches.
\item $y_S, y_T$: labels used in cross-entropy (\emph{CE}).
\item $\mathrm{CE}(\cdot,\cdot)$: cross-entropy classification loss between logits and labels (softmax over logits; averaged over the mini-batch).
\item $w_T(e)$: epoch-dependent weight on the target CE term.
\item $\lambda_{\mathrm{MMD}}(e)$: scale on the alignment term.
\item $\mathrm{MMD}^2(\mathrm{logits}_S,\mathrm{logits}_T)$: logit-level alignment penalty.
\end{itemize}

\subsubsection{Target-weighted supervised loss}
Let $\mathcal{D}_S=\{(x_i^S,y_i^S)\}_{i=1}^{n_S}$ and $\mathcal{D}_T=\{(x_j^T,y_j^T)\}_{j=1}^{n_T}$ be labeled source/target samples, and let $E$ be the number of epochs with current epoch $e\in\{1,\dots,E\}$. We use a warm-up factor
\begin{equation}
\alpha_e=\min\!\bigl(1,\, e/W\bigr)\in[0,1],
\label{eq:alpha}
\end{equation}
and define epoch-wise domain weights from the effective per-epoch counts $N_S,N_T$ as
\begin{IEEEeqnarray}{l}
w_{T0}=\sqrt{N_S/N_T},\\
\tilde w_T=\mathrm{clip}(w_{T0},1,6), \\
w_T=1+\alpha_e(\tilde w_T-1),\quad w_S=1.
\label{eq:weights}
\end{IEEEeqnarray}

where $\mathrm{clip}(x,a,b)=\min\{\max(x,a),\,b\}$.
With cross-entropy $\ell(\cdot,\cdot)$ and logits $f(\cdot;\theta)$, the supervised term is
\begin{align}
\mathcal{L}_{\text{sup}}
&= w_S\,\mathbb{E}_{(x,y)\sim\mathcal{D}_S}\!\big[\ell(f(x;\theta),y)\big]\nonumber\\
&+ w_T\,\mathbb{E}_{(x,y)\sim\mathcal{D}_T}\!\big[\ell(f(x;\theta),y)\big].
\label{eq:lsup}
\end{align}

\subsubsection{Logit-level RBF-MMD alignment}
For alignment, let the source/target \emph{logits} in a training step be $\mathcal{S}=\{s_i\}_{i=1}^{n}$ and $\mathcal{T}=\{t_j\}_{j=1}^{m}$. Using an RBF kernel
\begin{equation}
k(u,v)=\exp\!\Bigl(-\frac{\lVert u-v\rVert_2^2}{2\sigma^2}\Bigr),
\label{eq:kernel}
\end{equation}
with bandwidth $\sigma$ set by the median heuristic over $\mathcal{S}\cup\mathcal{T}$, the unbiased MMD is \cite{Gretton2012,Long2015DAN}
\begin{align}
\mathrm{MMD}^2(\mathcal{S},\mathcal{T})
&= \frac{1}{n(n-1)} \!\!\sum_{i\neq i'} k(s_i,s_{i'}) \nonumber\\
&+ \frac{1}{m(m-1)} \!\!\sum_{j\neq j'} k(t_j,t_{j'}) \nonumber\\
&- \frac{2}{nm}\sum_{i=1}^{n}\sum_{j=1}^{m} k(s_i,t_j).
\label{eq:mmd}
\end{align}
The full objective is
\begin{IEEEeqnarray}{l}
\mathcal{L}(e)=\mathcal{L}_{\text{sup}}+\lambda_{\mathrm{MMD}}(e)\,\mathrm{MMD}^2(\mathcal{S},\mathcal{T}), \\
\lambda_{\mathrm{MMD}}(e)=\alpha_e\,\lambda_0
\end{IEEEeqnarray}
which adds no trainable parameters beyond the global scale $\lambda_0$. Compared to layerwise MMD as in Deep Adaptation Networks (DAN) \cite{Long2015DAN}, aligning only logits keeps the method hyperparameter-light and computationally frugal for few-shot EEG.

\subsubsection{Split-BN}
Batch-normalization statistics are dataset-dependent; mixing domains in a single set of running moments can induce covariate shift at inference. Domain-aware normalization—e.g., AdaBN/DSBN—mitigates this by decoupling statistics across domains without duplicating the backbone \cite{Li2018AdaBN,Chang2019DSBN}. Let the network have BN layers indexed by $\ell\in\{1,\dots,L_{\mathrm{BN}}\}$. For each BN layer $\ell$, we share affine parameters $\gamma^{(\ell)},\beta^{(\ell)}$ across domains, but maintain separate running mean/variance buffers for each domain $d\in\{S,T\}$:
$
\mu_{d}^{(\ell)}, \sigma_{d}^{2(\ell)}.
$
Training alternates mini-batches from source and target. Before the forward pass of a mini-batch from domain $d$, we restore $\{\mu_{d}^{(\ell)},\sigma_{d}^{2(\ell)}\}_{\ell}$; after the pass, we snapshot the updated buffers. At evaluation on the target dataset, we always use the target buffers $\{\mu_{T}^{(\ell)},\sigma_{T}^{2(\ell)}\}_{\ell}$. In bidirectional experiments (AVO as source with P3 as target, and the reverse), each configuration uses its own target BN buffers for evaluation.

\subsection{Training Details}
\label{subsec:train}
We train the backbone with an adaptive optimizer and cosine-annealing scheduler. Mini-batch training is used with optional gradient accumulation. Regularization includes label smoothing, small temporal jitter, and additive Gaussian noise. Early stopping monitors the target-domain validation accuracy, and we select the checkpoint with the best target validation. Random seeds are set for all randomness sources to ensure reproducibility. Key hyperparameters are summarized in Table~\ref{tab:key_hparams}, and the training loop is detailed in Algorithm~\ref{alg:train}.

\renewcommand{\arraystretch}{1.3}
\begin{table}[H]
\caption{Key hyperparameters and implementation details}
\begin{center}
\begin{tabular}{|c|c|}
\hline
\textbf{Hyperparameter} & \textbf{Value} \\
\hline
Warm-up epochs $W$ & 40 \\
\hline
MMD base weight $\lambda_{0}$ & 0.4 (for $\sim$8:1 imbalance) \\
\hline
BN momentum & 0.1 \\
\hline
Optimizer & Adamax \\
\hline
Learning rate & 0.01 \\
\hline
Weight decay & $10^{-4}$ \\
\hline
Optimizer betas & (0.9, 0.999) \\
\hline
Batch size & 32 \\
\hline
Temporal jitter amplitude $\tau$ & $\pm5$ samples ($\approx$39 ms @ 128 Hz) \\
\hline
Gaussian noise std $\sigma_{\text{noise}}$ & 0.005 (z-scored units) \\
\hline
Early-stopping patience & 50 epochs \\
\hline
Early-stopping metric & Target-val accuracy \\
\hline
Braindecode version & 1.1.0 \\
\hline
Random seeds & $\{42,123,456,789,321\}$ \\
\hline
\end{tabular}
\label{tab:key_hparams}
\end{center}
\end{table}

\begin{algorithm}[H]
\caption{Training with Target-Focused Weighting, \mbox{Split-BN}, and Logits MMD}
\label{alg:train}
\begin{algorithmic}[1]
\FOR{epoch $e=1$ to $E$}
  \STATE $\alpha_e \leftarrow \min(1,\, e/W)$;\quad
         $w_T \leftarrow 1 + \alpha_e(\mathrm{clip}(\sqrt{N_S/N_T},1.0,6.0)-1)$;\quad
         $w_S \leftarrow 1$;\quad
         $\lambda_{\text{MMD}} \leftarrow \alpha_e\lambda_0$
  \FORALL{paired mini-batches $(\mathcal{B}_S,\mathcal{B}_T)$}
    \STATE \textbf{BN.use(S)}; \quad $z_S \leftarrow f(\mathcal{B}_S)$; \quad $L_S \leftarrow w_S\,\ell(z_S,y_S)$; \quad \textbf{BN.snapshot(S)}
    \STATE \textbf{BN.use(T)}; \quad $z_T \leftarrow f(\mathcal{B}_T)$; \quad $L_T \leftarrow w_T\,\ell(z_T,y_T)$; \quad \textbf{BN.snapshot(T)}
    \STATE $M \leftarrow \mathrm{MMD}^2(z_S, z_T)$ \hfill (RBF, median heuristic)
    \STATE $L \leftarrow L_S + L_T + \lambda_{\text{MMD}}\cdot M$; \quad backprop \& update
  \ENDFOR
\ENDFOR
\STATE \textbf{Inference on Target:} use BN(T) buffers.
\end{algorithmic}
\end{algorithm}

\subsection{Evaluation and confidence intervals}
\label{subsec:eval_ci}

All methods follow the protocol in Sec.~\ref{sec:datasets}: two source--target configurations, per-subject budgets (80 source and 10 target trials), and $k=5$-fold cross-validation repeated $r=5$ times, yielding $K=kr=25$ matched fold$\times$seed replicates per configuration. We pool trials across all subjects and perform stratified splits by trial. Thus, train/validation/test folds contain trials from all subjects (i.e., not leave-subject-out). This assesses generalization across trials within subjects rather than across subjects. 
For a metric $m$ (accuracy or AUC), let $\{m_i\}_{i=1}^{K}$ be scores on the target test split across replicates, with sample mean $\bar{m}$ and variance $s^2=\tfrac{1}{K-1}\sum_{i}(m_i-\bar m)^2$. We report the mean and the two-sided descriptive 95\% Student-$t$ interval:
\begin{equation}
\bar{m} \pm t_{0.975,\nu}\,\frac{s}{\sqrt{K}},\qquad \nu=K-1,
\label{eq:ci}
\end{equation}
For pairwise comparisons between two methods A and B, we compute per-replicate differences on the same fold$\times$seed split, $d_i=m_i^{(A)}-m_i^{(B)}$, then apply the corrected resampled paired $t$-test to account for dependence due to repeated cross-validation \cite{NadeauBengio2003,BouckaertFrank2004}. Define
\begin{IEEEeqnarray}{rCl}
\bar d &=& \tfrac{1}{K}\sum_{i=1}^{K} d_i, \label{eq:mean_diff}\\[2pt]
s_d^2 &=& \tfrac{1}{K-1}\sum_{i=1}^{K}(d_i-\bar d)^2, \label{eq:sd}\\[2pt]
\rho &=& \frac{n_{\mathrm{test}}}{n_{\mathrm{train}}}, \label{eq:rho}\\[2pt]
\gamma &=& \tfrac{1}{kr}+\rho. \label{eq:gamma}
\end{IEEEeqnarray}

where $n_{\mathrm{train}}$ and $n_{\mathrm{test}}$ are the target-domain train/test sizes per fold. The test statistic and degrees of freedom are
\begin{equation}
t_{\mathrm{corr}}=\frac{\bar d}{\sqrt{\gamma\, s_d^2}},\qquad \nu=K-1,
\label{eq:tcorr}
\end{equation}
from which we report the two-sided $p$-value.

\section{Results}\label{sec:results}
\begin{figure*}[!t]
  \centering
  \subfloat[Accuracy distributions on both datasets (baselines vs.\ AS-MMD). ]{%
    \includegraphics[width=0.48\textwidth]{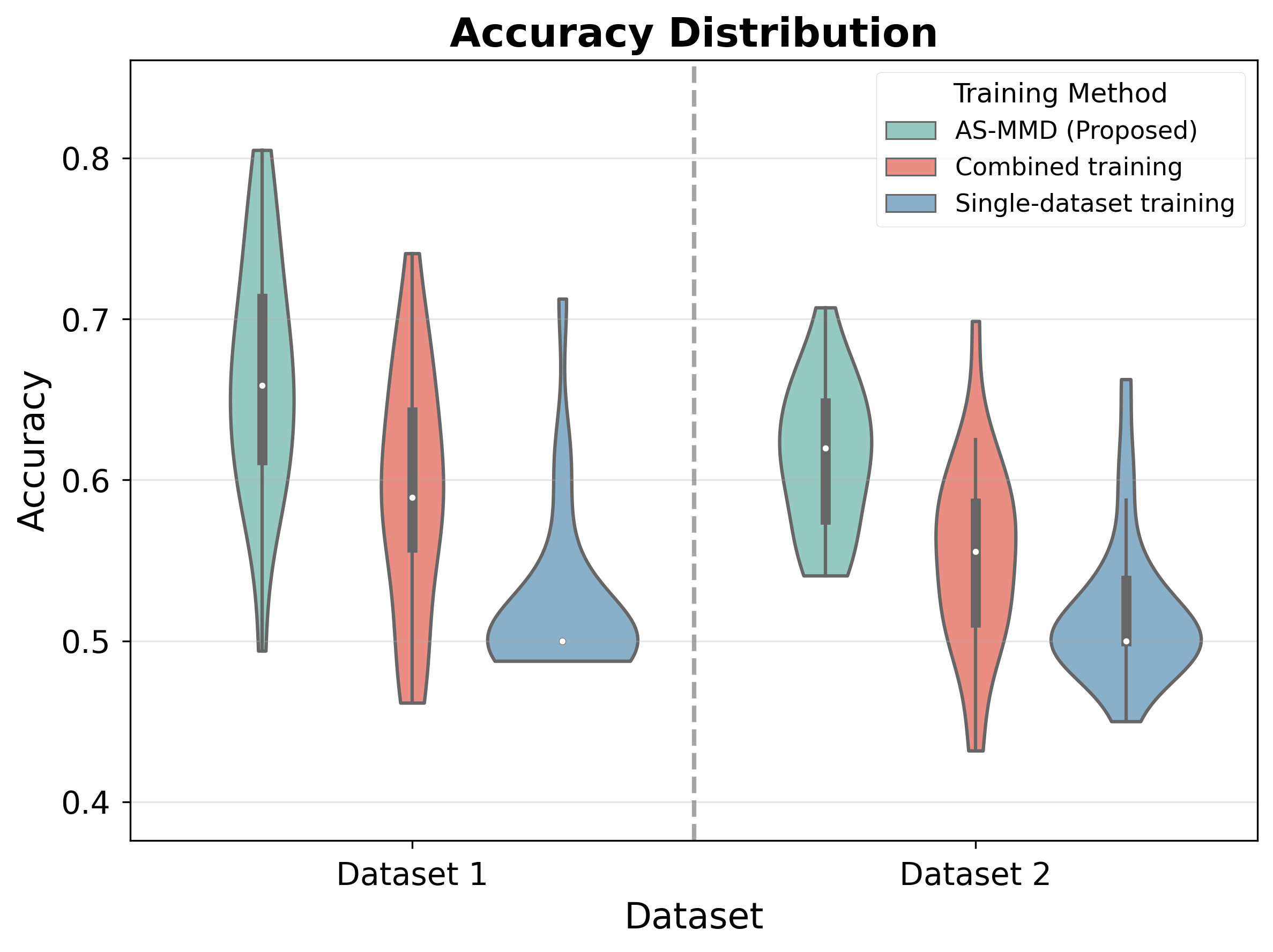}%
    \label{fig:violin_acc}}
  \hfil
  \subfloat[AUC distributions on both datasets (baselines vs.\ AS-MMD).]{%
    \includegraphics[width=0.48\textwidth]{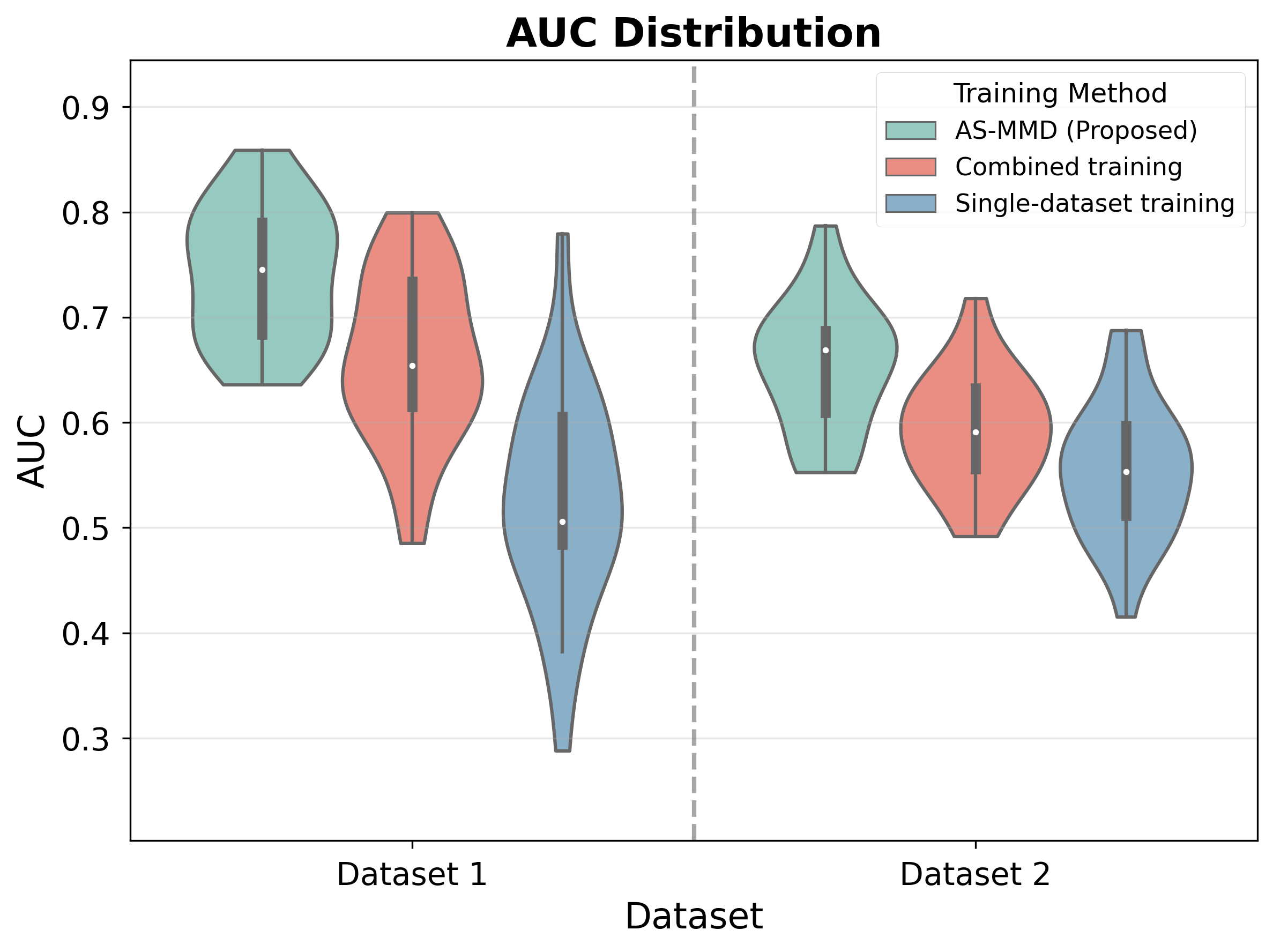}%
    \label{fig:violin_auc}}
  \caption{Comparison of baselines and AS-MMD on AVO and ERP CORE P3.}
\end{figure*}

Across both target datasets, AS-MMD consistently achieves higher AUC and accuracy than target-only and naive combined training, while exhibiting reduced run-to-run variability. Naive combining improves performance substantially on AVO and yields modest but statistically significant gains on ERP CORE P3; by contrast, AS-MMD provides consistent and larger improvements on both. Distributional evidence appears in Figs.~\ref{fig:violin_acc}--\ref{fig:violin_auc}, and numerical summaries with 95\% confidence intervals are provided in Table~\ref{tab:smallset_performance}.

\subsection{Comparison with Baselines}

Table~\ref{tab:smallset_performance} summarizes the classification performance on the two target small datasets (Dataset~1: Active Visual Oddball, AVO; Dataset~2: ERP CORE P3) under three training strategies: target-only training, combined training, and the proposed AS-MMD method. Results are reported in terms of mean accuracy and AUC, along with their 95\% confidence intervals across cross-validation folds.

\renewcommand{\arraystretch}{1.3}
\begin{table*}[htbp]
\caption{Performance on the two target datasets (means and 95\% CIs).}
\begin{center}
\begin{tabular}{|c|c|S[table-format=1.2]|c|S[table-format=1.2]|c|}
\hline
\textbf{Dataset} & \textbf{Training Method} & \textbf{Accuracy (Mean)} & \textbf{Accuracy (95\% CI)} & \textbf{AUC (Mean)} & \textbf{AUC (95\% CI)} \\
\hline
\multirow{3}{*}{Dataset 1: Active Visual Oddball}
& \textbf{AS-MMD (Proposed)} & {\bfseries 0.66} & {\bfseries {0.63--0.69}}& {\bfseries 0.74} & {\bfseries {0.72--0.77}} \\
\cline{2-6}
& Target-only training  & 0.52 & {0.50--0.54} & 0.53 & {0.48--0.57} \\
\cline{2-6}
& Combined training        & 0.59 & {0.56--0.62} & 0.66 & {0.63--0.69} \\
\hline
\multirow{3}{*}{Dataset 2: ERP CORE P3}
& \textbf{AS-MMD (Proposed) }       & {\bfseries 0.61} & {\bfseries {0.60--0.63}} & {\bfseries 0.65} & {\bfseries {0.63--0.68}} \\
\cline{2-6}
& Target-only training  & 0.52 & {0.50--0.54} & 0.56 & {0.53--0.58} \\
\cline{2-6}
& Combined training        & 0.55 & {0.53--0.58} & 0.59 & {0.57--0.62} \\
\hline
\end{tabular}
\label{tab:smallset_performance}
\end{center}
\end{table*}

On Dataset~1 (AVO), training on the small dataset alone yielded limited performance, with mean accuracy of 0.52 and AUC of 0.53. Incorporating the large dataset through combined training led to a clear and statistically significant improvement (accuracy: $p<0.001$, AUC: $p<0.001$), highlighting the benefit of additional cross-dataset information. The proposed AS-MMD method further improved both metrics to 0.66 accuracy and 0.74 AUC, with narrow confidence intervals. Compared with combined training, these gains were statistically significant (accuracy: $p=0.0031$, AUC: $p<0.001$), demonstrating the effectiveness of AS-MMD in leveraging the large dataset while prioritizing the deployment domain.

On Dataset~2 (ERP CORE P3), combined training improved over target-only training (accuracy 0.55 vs.\ 0.52; AUC 0.59 vs.\ 0.56), and these gains were statistically significant in our 5$\times$5 CV (accuracy: $p=0.011$, AUC: $p=0.0026$). AS-MMD further increased performance to 0.61 accuracy and 0.65 AUC, significantly outperforming combined training (accuracy: $p<0.001$, AUC: $p=0.0014$). 

These results indicate that while naive dataset merging yields modest but reliable gains on both datasets, AS-MMD delivers consistently larger improvements under cross-dataset shift.

\subsection{Ablation Studies}
\begin{figure*}[!t]
  \centering
  \subfloat[Accuracy distributions for AS\mbox{-}MMD and ablations on both datasets.]{%
    \includegraphics[width=0.48\textwidth]{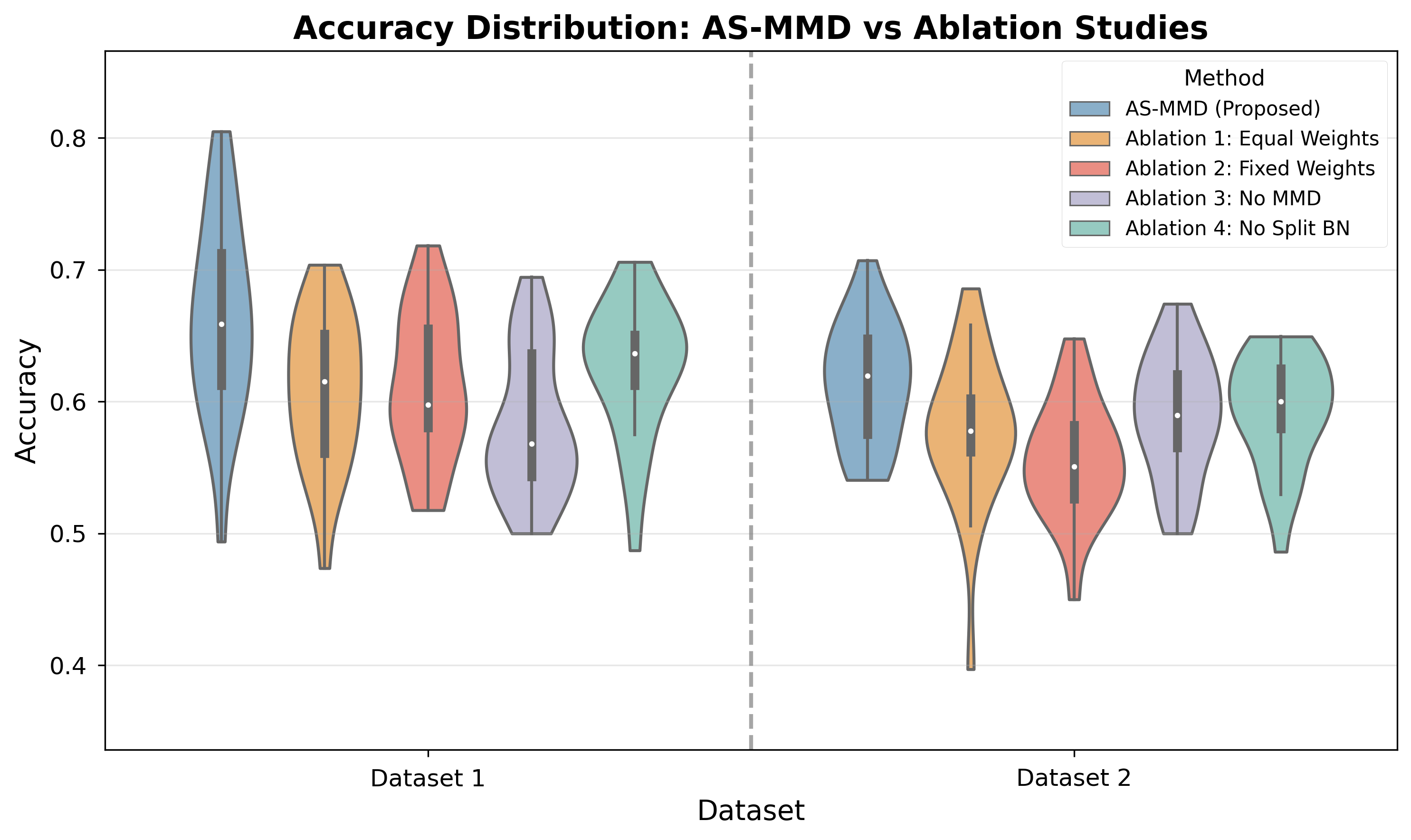}%
    \label{fig:ablation_violin_acc}}
  \hfil
  \subfloat[AUC distributions for AS\mbox{-}MMD and ablations on both datasets.]{%
    \includegraphics[width=0.48\textwidth]{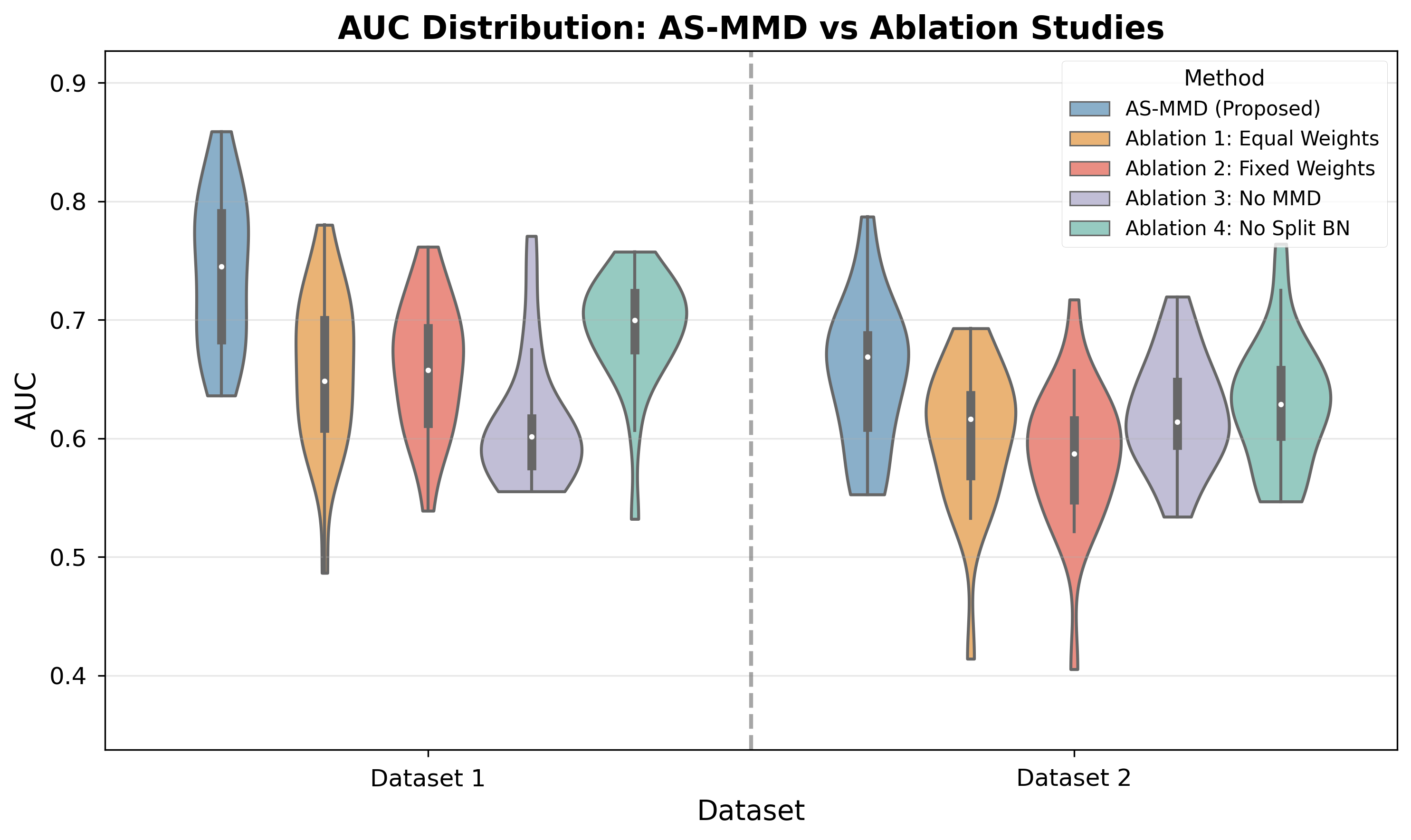}%
    \label{fig:ablation_violin_auc}}
  \caption{Ablation study: distributions of accuracy and AUC for AS\mbox{-}MMD and ablations on AVO and ERP CORE P3.}
\end{figure*}

To disentangle the contribution of each component, we conducted ablations of target-first weighting (Equal Weights, Fixed Weights), MMD alignment (No MMD), and Split-BN (No Split-BN). Distributional comparisons are shown in Figs.~\ref{fig:ablation_violin_acc}--\ref{fig:ablation_violin_auc}; detailed summaries appear in Table~\ref{tab:ablation_results}. The figures indicate that AS-MMD not only shifts the central tendency upward but also narrows dispersion relative to its ablations, particularly on AVO.

\renewcommand{\arraystretch}{1.3}
\begin{table*}[htbp]
\caption{Ablation results on the two target datasets (means and 95\% CIs).}
\begin{center}
\begin{tabular}{|c|c|S[table-format=1.2]|c|S[table-format=1.2]|c|}
\hline
\textbf{Dataset} & \textbf{Training Method} & \textbf{Accuracy (Mean)} & \textbf{Accuracy (95\% CI)} & \textbf{AUC (Mean)} & \textbf{AUC (95\% CI)} \\
\hline
\multirow{5}{*}{Dataset 1: Active Visual Oddball}
& \textbf{AS-MMD (Proposed)} & {\bfseries 0.66} & {\bfseries {0.63--0.69}} & {\bfseries 0.74} & {\bfseries {0.72--0.77}} \\
\cline{2-6}
& Equal Weights            & 0.64 & {0.63--0.65} & 0.73 & {0.72--0.73} \\
\cline{2-6}
& Fixed Weights            & 0.62 & {0.61--0.64} & 0.70 & {0.68--0.71} \\
\cline{2-6}
& No MMD                   & 0.60 & {0.58--0.62} & 0.69 & {0.68--0.70} \\
\cline{2-6}
& No Split-BN              & 0.65 & {0.64--0.66} & 0.72 & {0.71--0.73} \\
\hline
\multirow{5}{*}{Dataset 2: ERP CORE P3}
& \textbf{AS-MMD (Proposed) }   &   {\bfseries 0.61} & {\bfseries {0.60--0.63}} & {\bfseries 0.65} & {\bfseries {0.63--0.68}} \\
\cline{2-6}
& Equal Weights            & 0.58 & {0.55--0.60} & 0.60 & {0.58--0.63} \\
\cline{2-6}
& Fixed Weights            & 0.56 & {0.54--0.57} & 0.59 & {0.56--0.61} \\
\cline{2-6}
& No MMD                   & 0.59 & {0.57--0.61} & 0.62 & {0.60--0.64} \\
\cline{2-6}
& No Split-BN              & 0.59 & {0.58--0.61} & 0.63 & {0.61--0.65} \\
\hline
\end{tabular}
\label{tab:ablation_results}
\end{center}
\end{table*}

\paragraph*{Ablation t-tests}
On Dataset~1 (AVO), AS-MMD (Proposed) yielded statistically significant gains over all ablations in both metrics: Equal Weights (accuracy: $p = 0.0013$, AUC: $p < 0.001$), Fixed Weights (accuracy: $p = 0.016$, AUC: $p < 0.001$), No MMD (accuracy: $p < 0.001$, AUC: $p < 0.001$), and No Split-BN (accuracy: $p = 0.043$, AUC: $p = 0.0031$), indicating that each module—target-first weighting, Split-BN, and MMD alignment—contributes to the overall improvement.

On Dataset~2 (ERP CORE P3), AS-MMD (Proposed) showed statistically significant differences relative to most ablations: Equal Weights (accuracy: $p = 0.018$, AUC: $p = 0.012$), Fixed Weights (accuracy: $p < 0.001$, AUC: $p = 0.001$), and No MMD (accuracy: $p = 0.021$, AUC: $p = 0.017$). The contrasts with No Split-BN did not reach statistical significance (accuracy: $p = 0.11$, AUC: $p = 0.086$), suggesting that Split-BN plays a comparatively larger role on AVO than on ERP CORE P3.

The comprehensive results demonstrate that AS-MMD achieves higher average AUC and accuracy across both transfer directions. Without increasing inference complexity, it offers a low-cost, easily reusable training strategy, making it a suitable default starting point for small-sample cross-dataset EEG transfer learning.

\section{Discussion}
The findings indicate that a minimal, training-time modification yields clear relative gains over naive pooled training: on the AVO dataset, +11.86\% in accuracy and +12.12\% in AUC; on ERP CORE P3, +10.91\% in accuracy and +10.17\% in AUC. Relative to target-only training, the improvements are larger on AVO (+26.92\% accuracy; +39.62\% AUC) and remain consistent on ERP CORE P3 (+17.31\% accuracy; +16.07\% AUC). Beyond central performance, dispersion across folds and seeds narrows, suggesting improved stability under the strict small-sample target regime.

We interpret these outcomes through three training-time factors central to cross-dataset EEG. First, \emph{loss imbalance}: when source data dominate, the objective tilts toward the source distribution. Warm-started target weighting re-centers optimization on the deployment domain while leaving batch composition unchanged. Second, \emph{normalization drift}: mixing domains into a single set of batch statistics yields covariate shift at inference on the target domain; Split-BN avoids this by sharing affine parameters yet maintaining per-domain buffers, so evaluation on the target consistently uses target statistics. Third, \emph{residual decision-space mismatch}: even with supervised fitting, source and target logit distributions need not align; a parameter-free RBF–MMD term at the logit level provides a gentle, architecture-neutral correction.

Ablations clarify the role of each component without adding parameters or changing the inference procedure. Across datasets, the logit-level MMD provides the main improvement by reducing residual mismatch in the decision space after supervised fitting; its impact is most visible when the gap between source and target is larger. Target-focused weighting further improves performance by counteracting source dominance during optimization while leaving batch composition unchanged, which helps when naive pooling underserves the target objective. Split-BN contributes a smaller but reliable benefit by decoupling running statistics across domains and mitigating covariate shift at evaluation; the effect is dataset-dependent and can be modest when normalization drift is limited. Overall these mechanisms are complementary: alignment addresses output distribution mismatch, weighting recenters the objective on the deployment domain, and Split-BN stabilizes normalization, so the full recipe achieves higher central performance and lower variability than any single component alone.

Compared to layerwise alignment in Deep Adaptation Networks (DAN), which places MMD penalties on multiple hidden layers to progressively align representations \cite{Long2015DAN}, we deliberately impose a single, parameter-free penalty at the logit level. This keeps the method hyperparameter-light and computationally frugal. Extending AS-MMD toward layerwise or hybrid alignment in EEG (e.g., adding a weak MMD at the penultimate representation or adaptively weighting layers) is a promising direction for future work.

\paragraph*{Limitations}
Although AS-MMD improves target performance under the small-sample protocol, several study-specific factors qualify these gains. First, analyses were restricted to the set of electrodes shared across datasets, so the current recipe presumes overlapping montages and does not address heterogeneous electrode configurations. In addition, some contrasts did not reach statistical significance, indicating dataset-dependent sensitivity, particularly for components tied to normalization; therefore, the incremental benefit of Split-BN may be limited in certain regimes. Finally, our evaluation reflects generalization across trials within subjects rather than across subjects, which may lead to subject-specific fitting; this does not invalidate the findings, but future work should assess the method’s potential for across-subject generalization.

\section{Conclusion}
We introduced Adaptive Split-MMD Training (AS\mbox{-}MMD), a minimal, backbone\mbox{-}agnostic recipe for small\mbox{-}sample cross\mbox{-}dataset P300 decoding. By combining (i) a target\mbox{-}weighted loss with warm\mbox{-}up tied to the source/target size ratio, (ii) Split-BN with shared affine parameters and per\mbox{-}domain running statistics, and (iii) a parameter\mbox{-}free logit\mbox{-}level RBF-MMD term, AS\mbox{-}MMD improves single\mbox{-}trial oddball vs.\ standard classification over target\mbox{-}only and simply pooling on two public datasets using only five shared electrodes. Ablations confirm that each component contributes to the gains. The approach keeps the inference\mbox{-}time architecture unchanged, is simple to deploy, and provides a strong default for data\mbox{-}limited, cross\mbox{-}dataset EEG applications.


\bibliographystyle{IEEEtran}
\bibliography{refs}

\begin{thebibliography}{10}
\providecommand{\url}[1]{#1}
\csname url@samestyle\endcsname
\providecommand{\newblock}{\relax}
\providecommand{\bibinfo}[2]{#2}
\providecommand{\BIBentrySTDinterwordspacing}{\spaceskip=0pt\relax}
\providecommand{\BIBentryALTinterwordstretchfactor}{4}
\providecommand{\BIBentryALTinterwordspacing}{\spaceskip=\fontdimen2\font plus
\BIBentryALTinterwordstretchfactor\fontdimen3\font minus \fontdimen4\font\relax}
\providecommand{\BIBforeignlanguage}[2]{{%
\expandafter\ifx\csname l@#1\endcsname\relax
\typeout{** WARNING: IEEEtran.bst: No hyphenation pattern has been}%
\typeout{** loaded for the language `#1'. Using the pattern for}%
\typeout{** the default language instead.}%
\else
\language=\csname l@#1\endcsname
\fi
#2}}
\providecommand{\BIBdecl}{\relax}
\BIBdecl

\bibitem{HeWu2020EA}
H.~He and D.~Wu, ``Transfer learning for brain--computer interfaces: A euclidean space data alignment approach,'' \emph{IEEE Transactions on Biomedical Engineering}, vol.~67, no.~2, pp. 399--410, 2020.

\bibitem{Zhao2019}
Z.~Zhao, L.~Zhang, A.~Cichocki, and J.~Li, ``Robust transfer learning for motor imagery {EEG} classification,'' \emph{IEEE Transactions on Neural Systems and Rehabilitation Engineering}, vol.~27, no.~5, pp. 936--945, 2019.

\bibitem{Xu2021}
M.~Xu, F.~He, T.-P. Jung, X.~Gu, and D.~Ming, ``Spatio-temporal recurrent neural network for subject-independent erp detection,'' \emph{IEEE Transactions on Neural Systems and Rehabilitation Engineering}, vol.~29, pp. 1434--1445, 2021.

\bibitem{Jayaram2016}
V.~Jayaram, M.~Alamgir, Y.~Altun, B.~Scholkopf, and M.~Grosse-Wentrup, ``Transfer learning in brain-computer interfaces,'' \emph{IEEE Computational Intelligence Magazine}, vol.~11, no.~1, pp. 20--31, 2016.

\bibitem{Kostas2020ThinkerInvariance}
D.~Kostas and F.~Rudzicz, ``Thinker invariance: enabling deep neural networks for {BCI} across more people,'' \emph{Journal of Neural Engineering}, vol.~17, no.~5, p. 056008, 2020.

\bibitem{Song2023EEGConformer}
Y.~Song, Q.~Zheng, B.~Liu, and X.~Gao, ``{EEG} conformer: Convolutional transformer for {EEG} decoding and visualization,'' \emph{IEEE Transactions on Neural Systems and Rehabilitation Engineering}, vol.~31, pp. 710--719, 2023.

\bibitem{Isbell2025CESCA}
E.~Isbell, A.~N. Peters, D.~M. Richardson, and N.~E.~R. De~León, ``Cognitive electrophysiology in socioeconomic context in adulthood,'' \emph{Scientific Data}, vol.~12, p. 841, 2025.

\bibitem{Kappenman2021ERPCORE}
E.~S. Kappenman, J.~L. Farrens, W.~Zhang, A.~X. Stewart, and S.~J. Luck, ``{ERP} core: An open resource for human event-related potential research,'' \emph{NeuroImage}, vol. 225, p. 117465, 2021.

\bibitem{Polich2007P3Review}
J.~Polich, ``Updating p300: An integrative theory of p3a and p3b,'' \emph{Clinical Neurophysiology}, vol. 118, no.~10, pp. 2128--2148, 2007.

\bibitem{AlvaradoGonzalez2016P300}
M.~Alvarado-Gonz{\'a}lez, E.~Gardu{\~n}o, E.~Bribiesca, O.~Y{\'a}\~{n}ez Su{\'a}rez, and V.~Medina-Ba{\~n}uelos, ``P300 detection based on {EEG} shape features,'' \emph{Computational and Mathematical Methods in Medicine}, vol. 2016, pp. 1--14, 2016, article ID 2029791.

\bibitem{Lawhern2018EEGNet}
V.~J. Lawhern, A.~J. Solon, N.~R. Waytowich, S.~M. Gordon, C.~P. Hung, and B.~J. Lance, ``{EEG}net: A compact convolutional neural network for {EEG}-based brain--computer interfaces,'' \emph{Journal of Neural Engineering}, vol.~15, no.~5, p. 056013, 2018.

\bibitem{Schirrmeister2017HBM}
R.~T. Schirrmeister, J.~T. Springenberg, L.~D.~J. Fiederer, M.~Glasstetter, K.~Eggensperger, M.~Tangermann, F.~Hutter, W.~Burgard, and T.~Ball, ``Deep learning with convolutional neural networks for {EEG} decoding and visualization,'' \emph{Human Brain Mapping}, vol.~38, no.~11, pp. 5391--5420, 2017.

\bibitem{Xu2020CrossDataset}
L.~Xu, M.~Xu, Y.~Ke, X.~An, S.~Liu, and D.~Ming, ``Cross-dataset variability problem in {EEG} decoding with deep learning,'' \emph{Frontiers in Human Neuroscience}, vol.~14, p. 103, 2020.

\bibitem{Zanini2018RA}
P.~Zanini, M.~Congedo, C.~Jutten, S.~Said, and Y.~Berthoumieu, ``Transfer learning: A riemannian geometry framework with applications to brain--computer interfaces,'' \emph{IEEE Transactions on Biomedical Engineering}, vol.~65, no.~5, pp. 1107--1116, 2018.

\bibitem{Li2018AdaBN}
Y.~Li, N.~Wang, J.~Shi, J.~Liu, and X.~Hou, ``Adaptive batch normalization for practical domain adaptation,'' \emph{Pattern Recognition}, vol.~80, pp. 109--117, 2018.

\bibitem{Chang2019DSBN}
W.-G. Chang, T.~You, S.~Seo, S.~Kwak, and B.~Han, ``Domain-specific batch normalization for unsupervised domain adaptation,'' in \emph{Proc. IEEE/CVF Conf. on Computer Vision and Pattern Recognition (CVPR)}, 2019.

\bibitem{Kilani2024SourceSel}
S.~Kilani, S.~N. Aghili, Y.~Fathi, and A.~I. Sburlea, ``Optimization of transfer learning based on source sample selection in euclidean space for p300-based brain--computer interfaces,'' \emph{Frontiers in Neuroscience}, vol.~18, p. 1360709, 2024.

\bibitem{Long2015DAN}
M.~Long, Y.~Cao, J.~Wang, and M.~I. Jordan, ``Learning transferable features with deep adaptation networks,'' in \emph{Proc. Int. Conf. on Machine Learning (ICML)}, 2015, pp. 97--105.

\bibitem{Gramfort2013MNEPython}
A.~Gramfort, M.~Luessi, E.~Larson, D.~Engemann, D.~Strohmeier, C.~Brodbeck, R.~Goj, M.~Jas, T.~Brooks, L.~Parkkonen, and M.~H{\"a}m{\"a}l{\"a}inen, ``{MEG} and {EEG} data analysis with {MNE}-python,'' \emph{Frontiers in Neuroscience}, vol.~7, p. 267, 2013.

\bibitem{Gramfort2014MNE}
A.~Gramfort, M.~Luessi, E.~Larson, D.~A. Engemann, D.~Strohmeier, C.~Brodbeck, L.~Parkkonen, and M.~S. H{\"a}m{\"a}l{\"a}inen, ``{MNE} software for processing {MEG} and {EEG} data,'' \emph{NeuroImage}, vol.~86, pp. 446--460, 2014.

\bibitem{Jung2000ICA}
T.-P. Jung, S.~Makeig, C.~Humphries, T.-W. Lee, M.~J. McKeown, V.~Iragui, and T.~J. Sejnowski, ``Removing electroencephalographic artifacts by blind source separation,'' \emph{Psychophysiology}, vol.~37, no.~2, pp. 163--178, 2000.

\bibitem{Gretton2012}
A.~Gretton, K.~M. Borgwardt, M.~J. Rasch, B.~Schölkopf, and A.~Smola, ``A kernel two-sample test,'' \emph{Journal of Machine Learning Research}, vol.~13, pp. 723--773, 2012.

\bibitem{NadeauBengio2003}
C.~Nadeau and Y.~Bengio, ``Inference for the generalization error,'' \emph{Machine Learning}, vol.~52, pp. 239--281, 2003.

\bibitem{BouckaertFrank2004}
R.~R. Bouckaert and E.~Frank, ``Evaluating the replicability of significance tests for comparing learning algorithms,'' in \emph{Proc. Pacific-Asia Conf. Knowl. Discov. Data Mining}, 2004, pp. 3--12.

\end{thebibliography}

\end{document}